\documentclass[aps,prb,twocolumn,footinbib,superscriptaddress,longbibliography]{revtex4-2}
\usepackage[draft]{graphicx}
\usepackage{bm}
\usepackage{amsmath,amssymb}
\allowdisplaybreaks
\usepackage[bookmarksnumbered,bookmarksopen]{hyperref}
\hypersetup{
   colorlinks=true,
   citecolor=blue,
}

\usepackage{braket}
\usepackage{array}
\newcolumntype {s}[1]{@{\hspace*{#1}}}
\newcommand* {\tvek}[2][c]{\left( \begin{array}{s{0.15em}#1s{0.15em}}
     #2\end{array} \right)}

\newcommand* {\vek}[1]{{\bm{\mathrm{#1}}}}
\newcommand* {\pp}{\vek{p}}

\begin{document}

\title{Time inversion symmetry in the Dirac and Schr\"odinger-Pauli theories}

\author{R. Winkler}
\affiliation{Department of Physics, Northern Illinois University,
DeKalb, Illinois 60115, USA}
\affiliation{Materials Science Division, Argonne National Laboratory,
Argonne, Illinois 60439, USA}

\author{U. Z\"ulicke}
\affiliation{MacDiarmid Institute, School of Chemical and Physical
Sciences, Victoria University of Wellington, PO Box 600, Wellington
6140, New Zealand}

\date{August 12, 2025}

\begin{abstract}
The Schr\"odinger-Pauli theory is generally believed to give a
faithful representation of the nonrelativistic and weakly
relativistic limit of the Dirac theory.  However, the
Schr\"odinger-Pauli theory is fundamentally incomplete in its
account of broken time inversion symmetry, e.g., in magnetically
ordered systems.  In the Dirac theory of the electron, magnetic
order breaks time inversion symmetry even in the nonrelativistic
limit, whereas time inversion symmetry is effectively preserved in
the Schr\"odinger-Pauli theory in the absence of spin-orbit
coupling.  In the Dirac theory, the Berry curvature $1/(2m^2c^2)$ is
thus an intrinsic property of nonrelativistic electrons similar to
the well-known spin magnetic moment $e\hbar/(2m)$, while this result
is missed by the nonrelativistic or weakly relativistic
Schr\"odinger-Pauli equation.  In ferromagnetically ordered systems,
the intrinsic Berry curvature yields a contribution to the anomalous
Hall conductivity independent of spin-orbit coupling.
\end{abstract}

\maketitle

While the famous Dirac equation provides a relativistically
invariant formulation of quantum mechanics, the Schr\"odinger-Pauli
equation with a Zeeman term and a spin-orbit coupling (SOC) term is
generally believed to properly represent the nonrelativistic or
weakly relativistic limit of the Dirac equation \cite{str98}.  Time
inversion symmetry (TIS) is a fundamental symmetry of nature that is
broken by magnetic order.  Broken TIS represents the fundamental
cause for all physical phenomena that distinguish magnetic systems
from nonmagnetic systems \cite{lan8e}.  This includes, e.g., the
anomalous Hall effect \cite{nag10} and the magnetoelectric effect
\cite{lan8e}.

This Letter demonstrates that the Dirac theory accounts for the
breaking of TIS in magnetically ordered structures even in the
nonrelativistic limit and without SOC.  The distinction between
systems preserving TIS and systems breaking TIS is lost in the
nonrelativistic Schr\"odinger theory; and it can only be
re-introduced in the weakly relativistic Pauli theory via SOC.  Our
findings are relevant in the context of recent Schr\"odinger-Pauli
theories of magnetic order distinguishing between nonrelativistic
magnetic phenomena arising in the absence of SOC and phenomena that
do require SOC \cite{sme22, sme22a}.

For conceptual simplicity, we use a 2D model to illustrate the
qualitative differences between the Dirac theory and
Schr\"odinger-Pauli theory when applied to nonrelativistic magnetic
electron systems.  Ignoring the $z$ component of motion, the $4
\times 4$ Dirac Hamiltonian becomes
\begin{equation}
  \label{eq:dirac}
  H_{4\times 4} = \tvek[cccc]{
  \Delta^+ & 0 & 0 & c p_- \\
  0 & \Delta^- & c p_+ & 0 \\
  0 & c p_- & - \Delta^- & 0 \\
  c p_+ & 0 & 0 & - \Delta^+} \, ,
\end{equation}
where $p_\pm \equiv p_x \pm ip_y$ and $\Delta^\pm = mc^2 \pm
\delta$. The parameter $\delta \ll mc^2$ represents a simple model
for an exchange field arising from magnetic order.  We ignore any
potential $V$.  By regrouping rows and columns appropriately, the
Hamiltonian (\ref{eq:dirac}) can be rewritten in unitarily
equivalent, block-diagonal form as
\begin{subequations}
  \begin{equation}
    \tilde{H}_{4\times 4}
    = \tvek[cc]{H_{2\times 2}^+ & 0 \\ 0 & H_{2\times 2}^-}
  \end{equation}
  with
  \begin{equation}
    H_{2\times 2}^\pm =
    \tvek[cc]{\Delta^\pm & c p_\mp \\ c p_\pm & -\Delta^\pm} \, .
  \end{equation}
\end{subequations}
Here $H_{2\times 2}^\pm$ are decoupled $2\times 2$ Hamiltonians for
spin up and spin down.  In the limit $\delta \rightarrow 0$, we have
$(H_{2\times 2}^\pm)^\ast = H_{2\times 2}^\mp \ne H_{2\times
2}^\pm$, i.e., the decoupled Hamiltonians are inherently complex,
indicating that the Hamiltonians $H_{2\times 2}^\pm$ individually
\emph{always} break TIS.  Thus, very generally, the (backward)
unitary time evolution under $(H_{2\times 2}^\pm)^\ast$ cannot annul
the (forward) unitary time evolution under $H_{2\times 2}^\pm$, as
it is the case in systems preserving TIS.

Unlike the Schr\"odinger-Pauli theory, the Dirac theory does not
explicitly contain a magnetic moment of the electron.  But the
electrons' orbital motion gives rise to an orbital magnetic moment
of the eigenstates of $H_{2\times 2}^\pm$ that can be evaluated with
the modern theory of orbital magnetization \cite{xia07, res10,
misc:cross}, yielding (here for the positive-energy eigenstates
$\ket{\psi^\pm}$)
\begin{subequations}
  \begin{align}
    \mu^\pm (p) & = \pm \frac{ie\hbar}{2} \,
    \Braket{\vek{\nabla}_\pp \psi^\pm |
    \times \left[ H_{2\times 2}^\pm - E^\pm(\pp) \right]
    | \vek{\nabla}_\pp \psi^\pm}_z   \label{eq:magnet:def} \\*
                & = \pm \frac{e\hbar \, c^2 \Delta^\pm}
    {2 \bigl({\Delta^\pm}^2 + c^2 p^2\bigr)}
    \approx \pm \frac{e\hbar \, mc^4}
    {2 \left(m^2 c^4 + c^2 p^2\right)} \, .
\end{align}
\end{subequations}
As expected, the $p \rightarrow 0$ limit of the orbital magnetic
moment in the Dirac theory equals the well-known spin magnetic
moment
\begin{equation}
  \label{eq:spinmag}
  \mu_s^\pm = \pm \frac{e\hbar}{2m}
\end{equation}
in the nonrelativistic Schr\"odinger-Pauli theory \cite{misc:shi07,
misc:win20}.  Similarly, we can evaluate for the eigenstates
$\ket{\psi^\pm}$ of $H_{2\times 2}^\pm$ the Berry curvature
\cite{xia07, cul03, misc:cross}
\begin{subequations}
  \label{eq:berry}
  \begin{align}
  \Omega^\pm (p) & = i \, \Braket{ \vek{\nabla}_\pp \psi^\pm |
    \times | \vek{\nabla}_\pp \psi^\pm}_z   \label{eq:berry:def} \\*
  & = \pm \frac{c^2 \Delta^\pm}
  {2 \bigl({\Delta^\pm}^2 + c^2 p^2\bigr)^{3/2}}
  \approx \pm \frac{mc^4}
  {2 \left(m^2 c^4 + c^2 p^2\right)^{3/2}}
\end{align}
\end{subequations}
that determines the anomalous Hall conductivity \cite{nag10}.  In the
nonrelativistic limit $p \rightarrow 0$, we obtain \cite{cha08}
\begin{equation}
  \label{eq:berry0}
  \Omega^\pm (0) = \pm \frac{c^2}{2 {\Delta^\pm}^2}
  \approx \pm \frac{1}{2 m^2 c^2} \, .
\end{equation}
The Berry curvature (\ref{eq:berry0}) is an intrinsic property of
the nonrelativistic electron like the famous spin magnetic moment
(\ref{eq:spinmag}).
An imbalance $\Delta n$ between spin-up and -down states in a
ferromagnet thus implies, besides a net orbital magnetic moment, a
net intrinsic anomalous Hall conductivity $\sim (e^2/\hbar) (\hbar /
m c)^2 \Delta n /2$.  Using typical numbers, this simple estimate
yields conductivities that are too small to explain experimentally
observed values, but it illustrates the fundamental difference
between models preserving TIS and models breaking TIS.  Different
from earlier work \cite{nag10}, SOC due to a potential $V$ is not
necessary for this contribution to the anomalous Hall conductivity.

In the nonrelativistic Schr\"odinger theory, the Hamiltonians
$H_{2\times 2}^\pm$ are replaced by
\begin{equation}
  \label{eq:ham-1}
  H_{1\times 1}^\pm = \frac{p^2}{2m} \pm \delta \, .
\end{equation}
These Hamiltonians are real, $(H_{1\times 1}^\pm)^\ast = H_{1\times
1}^\pm$, i.e., they preserve TIS individually even in the presence
of an exchange-field $\delta$---as noted previously in the context
of electronic-structure calculations for magnetic systems in, e.g.,
Refs.~\cite{gos15, yua20, sme22}; see also Ref.~\cite{bih07}
reviewing the general context.  This is closely related to the
well-known fact that nondegenerate eigenfunctions of Hamiltonians
$H_{1\times 1}^\pm$ are real upto an overall phase factor.  Beyond
that, TIS of spin-decoupled Schr\"odinger-type \mbox{$1 \times 1$}
crystal Hamiltonians $\mathcal{H}_{1\times 1}^\pm$ \cite{bih07}
implies that the Bloch eigenfunctions of $\mathcal{H}_{1\times
1}^\pm$ can be classified according to Herring's cases (a), (b), and
(c) \cite{her37}, which is a unique property of crystal Hamiltonians
preserving TIS.  This happens even if these decoupled Hamiltonians
$\mathcal{H}_{1\times 1}^\pm$ ultimately represent opposite spin
channels in (collinear) magnetically ordered structures.  Therefore,
such Hamiltonians cannot account for magnetic phenomena like the
Berry curvature~(\ref{eq:berry}) and the associated anomalous Hall
conductivity.
A nonzero Berry curvature representing the breaking of TIS is
usually re-introduced into the weakly relativistic Pauli theory via
SOC (arising from gradients of a potential $V$) that represents an
off-diagonal coupling between the blocks $H_{1\times 1}^\pm$ or
$\mathcal{H}_{1\times 1}^\pm$ \cite{nag10}.  Unlike the intrinsic
Berry curvature~(\ref{eq:berry}), the curvature in the Pauli theory
thus depends on the potential~$V$.

Similar to the anomalous Hall effect, it has previously been assumed
that microscopic theories of the magneto\-electric effect require SOC
\cite{don19}.  The magnetoelectric effect exists in materials that
break both TIS and space inversion symmetry \cite{lan8e}.  In fact,
such microscopic theories can again be developed independent of SOC
as long as the broken inversion symmetries are otherwise taken into
account.  However, in this case, the necessary details are more
intricate \cite{win20}, and a full discussion of these will be presented
elsewhere.

In conclusion, a proper account of magnetic phenomena is naturally
achieved in electronic-structure calculations based on the Dirac
theory where TIS breaking manifests itself via orbital equilibrium
currents.  Such fully relativistic calculations of magnetic systems,
though less common than calculations based on the
Schr\"odinger-Pauli theory, have been reported, e.g., in
Refs.~\cite{low10, wag19}.  However, such a description does not
lend itself to a decoupling of real-space order and magnetic order
that is assumed, e.g., in applications of spin-group theories
\cite{sme22, sme22a, lit74}.

\begin{acknowledgments}
  RW and UZ acknowledge discussions with D.~Culcer, T.~Jungwirth,
  J.~Sinova, and L.~\v{S}mejkal.  RW also acknowledges discussions
  with G.~Bihlmayer and D.~Vanderbilt.  Work at Argonne was
  supported by DOE BES under Contract No.\ DE-AC02-06CH11357.
\end{acknowledgments}


\begin{thebibliography}{21}
\bibitem [{misc:cross()}]{misc:cross}%
  \BibitemOpen
  \bibinfo {note} {In Eqs.\ (\ref{eq:magnet:def}) and (\ref{eq:berry:def}), the
  bra and ket are three-component vectors, and the symbol ``$\times$'' denotes
  the vector product of such vectors.}\BibitemShut {Stop}%
\end{thebibliography}
\end{document}